\documentclass[10pt, twocolumn]{IEEEtran}
\usepackage{graphicx,multirow}
\usepackage{epsfig}
\usepackage{algorithm}
\usepackage{algorithmic}
\usepackage{cite}
\usepackage{amssymb}
\usepackage{amsthm}
\usepackage{amsmath}
\usepackage{color}
\usepackage{bbm}
\usepackage{cite}
\usepackage{epstopdf}
\usepackage[table,xcdraw]{xcolor}
\newtheorem{theorem}{\textbf{\text{Theorem}}}

\usepackage{subcaption}
\usepackage{collcell, datatool}
\usepackage{pifont}

\definecolor{lightgray}{gray}{0.9}
\usepackage{algorithm} 
\usepackage{algorithmic}  
\usepackage[algo2e]{algorithm2e} 

\begin{document}

\title{
Tractable Stochastic Geometry Model for IoT Access in LTE Networks}
\author{
\IEEEauthorblockN{\large  Mohammad Gharbieh, Hesham ElSawy, Ahmed Bader, and Mohamed-Slim Alouini\\
\IEEEauthorblockA{\small  King Abdullah University of Science and Technology (KAUST), \\
Computer, Electrical, and Mathematical Sciences and Engineering (CEMSE) Divison, Thuwal, Makkah Province, Saudi Arabia,\\  
Email: \{mohammad.gharbieh, hesham.elsawy, ahmed.bader, slim.alouini\}@kaust.edu.sa}
}\vspace{-1.0cm}}

\maketitle
\thispagestyle{empty}
\pagestyle{empty}
\begin{abstract}

The Internet of Things (IoT) is large-scale by nature. This is not only manifested by the large number of connected devices, but also by the high volumes of traffic that must be accommodated. Cellular networks are indeed a natural candidate for the data tsunami the IoT is expected to generate in conjunction with legacy human-type traffic. However, the random access process for scheduling request represents a major bottleneck to support IoT via LTE cellular networks. Accordingly, this paper develops a mathematical framework to model and study the random access channel (RACH) scalability to accommodate IoT traffic. The developed model is based on stochastic geometry and discrete time Markov chains (DTMC) to account for different access strategies and possible sources of inter-cell and intra-cell interferences. To this end, the developed model is utilized to assess and compare three different access strategies, which incorporate a combination of transmission persistency, back-off, and power ramping. The analysis and the results showcased herewith clearly illustrate the vulnerability of the random access procedure as the IoT intensity grows. Finally, the paper offers insights into effective scenarios for each transmission strategy in terms of IoT intensity and RACH detection thresholds.


\end{abstract}
 \vspace{-5mm}

\section{Introduction}

The Internet of Things (IoT) is expected to involve a massive number of sensors, smart physical objects, machines, vehicles, and devices that communicate together and/or connect to the Internet \cite{IOT1}. Based on the IoT concept, a plethora of emerging applications are being proposed including vehicular communication, proximity services, autonomous driving, public safety, massive sensors support, and smart cities applications~ \cite{IOT1}.  However, the last mile wireless access represents a fundamental challenge and a limiting performance obstacle to realize IoT applications, especially for applications that involve mobility. In this context, cellular networks stand out of all other alternatives as a reliable, efficient, and ubiquitous radio access network (RAN) to provide IoT last mile connectivity~\cite{IOT2}. Consequently, the next evolution of cellular networks is not only envisioned to offer tangible performance improvement in terms of data rate, network capacity, energy efficiency, and latency, but also to support IoT applications. In addition to serving the legacy users, the cellular network should provide occasional Internet access for massive number of connected things. In other words, the cellular infrastructure should be able to accommodate unprecedented traffic levels that are essentially a blend of human-type and machine-type communications.

 Although each IoT element (i.e., thing) may have low traffic profile, the aggregate traffic generated from the IoT can be overwhelming \cite{IOT2, RACH1, RACH2}. As a matter of fact, when the frame inter-arrival time is large, the random access procedure is typically invoked twice for each uplink frame to be transmitted \cite{sesia2009lte, RACH2}. The first corresponds to the transition from idle (RRC\_IDLE) state to the connected (RRC\_CONNECTED) state. The second step is associated with the need of the device\footnote{Throughout this paper user equipment (UEs) and IoT elements are referred to as devices.} to send a scheduling request (SR) to the base station \cite{sesia2009lte, RACH2}. While some high-priority devices may be granted permission to send SRs on given uplink resources, the vast majority of devices are not synchronized and have to  contend on the RACH to request UL resources. This is specifically true when the number of devices is quite large. While synchronized devices encounter one random access process for SR, unsynchronized devices encounter two random access processes for synchronization and SR. This paper is concerned with the study of the success of the random access procedure irrespective of the actual state the device may be in.

 The scalability of the LTE to accommodate the massive RACH signaling, imposed by the IoT, via its current settings is questionable. For instance, \cite{RACH1, RACH2} show that the default LTE RACH access fails to support different IoT scenarios. However, the studies in \cite{RACH1, RACH2} for LTE RACH performance for IoT applications are confined to computationally complex simulations. Therefore, there is an urge need to develop a mathematical framework that parametrizes the RACH performance in terms the network parameters, traffic intensity, and IoT intensity. Such mathematical model is necessary to understand the LTE random access behavior, when the devices intensity scales, in order to pinpoint bottlenecks and draw legitimate conclusions about the RACH performance. In this context, stochastic geometry can be exploited to develop rigorous mathematical frameworks to conduct such scalability studies in the context of IoT.  Stochastic geometry is powerful mathematical tool that is able to incorporate large-scale spatial randomness, which is intrinsic in IoT, along with other sources of uncertainties that emerge in wireless networks into tractable analysis~\cite{survey_h, martin_book}.


By virtue of stochastic geometry, several models are developed to characterize the performance of cellular networks, see~\cite{survey_h} for a survey. 
However, the RACH performance of LTE is not yet modeled, especially in the presence of massive number of access attempts, as in the case of IoT. Note that the uplink normal data transmission models that exist in the literature (e.g., \cite{elsawy2014stochastic, uplink2_jeff})  cannot be directly generalized to capture the RACH access performance in IoT environments for three reasons. First,  the uplink data transmission is coordinated via the base station (BS) such that no intra-cell interference exist. On the other hand, the RACH channel access is uncoordinated and random, which may lead to intra-cell interference in addition to the inter-cell interference. Second, the RACH access scheme has different power control and back-off states that are not present in the regular data transmission mode. Finally, the massive number of simultaneous access attempts, that may take place in IoT scenarios, may lead to inter-cell interference with multiple interferers per BS.

This paper develops a mathematical model, based on stochastic geometry and discrete time Markov chains (DTMC), for LTE RACH access performance for IoT applications. While stochastic geometry accounts for the spatial intra and inter-cell sources of interference, the DTMC models the different RACH access schemes that are used by the devices. Particularly, we model three different types of RACH access schemes that offer different tradeoffs between transmission persistency, random back-off coordination, and power ramping. The main performance metrics considered are the ``{\em {\rm RACH} transmission failure probability} and the ``{\em average waiting time for {\rm RACH} success}". The developed model is then used to assess and compare the performance of the aforementioned RACH access schemes, which are defined by the LTE standard~\cite{sesia2009lte}. To the best knowledge of the authors, this is the first paper to develop a mathematical model for the LTE RACH access for IoT  applications in large scale environment. 
 
The results show that each RACH scheme has its own effective operation scenario to minimize the average waiting time for RACH access. At low device intensity, the average time for RACH access is minimized via the baseline scheme and power ramping scheme at, respectively, low and high signal-to-interference-plus-noise-ratio (SINR) thresholds. Particularly, at $0$ dB SINR threshold and intensity of $64$ device/BS, the power ramping technique reduces the average waiting time by $56\%$ when compared to the back-off scheme.  Which shows that back-off scheme imposes unnecessary delay in case of low device intensity.  As the intensity of devices starts to grow, prioritizing devices which encounter failures with power ramping is sufficient to minimize the average waiting time at moderate intensity and moderate SINR thresholds. However, the back-off scheme becomes crucial as the intensity or SINR threshold scales. That is, back-off becomes necessary to relief RACH congestion, maintains acceptable RACH transmission success probability, and hence, minimize the average waiting time for RACH success. For instance, the back-off scheme shows a reduction of $65\%$ and $99\%$ in the average waiting time for RACH success at 0 dB SINR threshold when compared to the power ramping scheme at $256$ devices/BS and $512$ devices/BS, respectively. It is worth mentioning that the results are obtained for BS intensity of 3 BSs/km$^2$ assuming the typical 64 orthogonal RACH sequences per BS. Hence, the aforementioned $64$,  $256$, and $512$ devices/BS intensities correspond to $192$,  $768$, and $1536$ devices/km$^2$, respectively. 


\section{System Model and Assumptions}\label{System Model}

\subsection{Network and Propagation Models}
 We consider a single-tier cellular network where the BSs are spatially distributed in $\mathbb{R}^2$ according to a homogeneous Poisson point process (PPP) $\Psi=\{x_k;k=1,2,3,.... \}$ with intensity $\lambda$. The devices are spatially distributed in $\mathbb{R}^2$ via an independent PPP  $\Phi=\{u_i;i=1,2,3,.... \}$ with intensity $\mathcal{U}$. 
 Without loss of generality, all BSs are assumed to have an open access policy, and hence, each of the devices is assumed to request Internet access from its nearest BS. 
 
 A general power-law path-loss model is considered where the signal power decays at a rate $r^{-\eta}$ with the propagation distance $r$, where $\eta>2$ is the path-loss exponent. In addition to the path-loss attenuation, all the channel gains are assumed to be independent of each other, independent of the spatial locations, and are identically distributed (i.i.d). For analysis, Rayleigh fading is considered, and hence, the channel power gains ($h$) are exponentially distributed and with unity mean.
 
 \subsection{RACH Access Scheme}
 
  To request channel access, each device randomly and independently transmits its request on one of the available prime-length Zadoff-Chu (ZC) sequences defined by the LTE physical random access channel (PRACH) preamble \cite{sesia2009lte}. It is assumed that the intensity of the IoT is high enough such that there are multiple active devices in each BS using the same Zadoff-Chu (ZC) sequence to request resource allocations \cite{sesia2009lte}. Without loss of generality, we assume that all BSs have the same number of ZC sequences, different ZC codes are orthogonal\footnote{That is, the BSs are dense enough such that all the sequences are generated from cyclic shifts of a single root sequence.}, and that the devices interfering on the same ZC code constitute a PPP $ \tilde{\Phi} \subseteq \Phi$ with intensity $\tilde{\mathcal{U}}=\frac{ \mathcal{T} \mathcal{U}}{{n_{Z} }}$, where $\mathcal{T}$ is the probability of transmission and  $n_{ Z}$ is the number of available ZC sequences.
 
All of the devices use full inversion power control with threshold $\rho$. That is, each device controls its transmit power such that the average signal power received at the corresponding serving BS is equal to a predefined power value $\rho$, which is assumed to be the same for all the BSs. It is assumed the BSs are dense enough such that each of the devices can invert its path loss towards the closest BS almost surely, so the maximum transmit power of the UEs is not a binding constraint for the RACH access. Extension to RACH access with fractional power control and adding a maximum power constraint can be done by following the methodologies in \cite{uplink2_jeff} and \cite{uplink_alamouri}, respectively. Upon RACH access failure, the ZC code selection is
repeated and the device follow one of the following three schemes:
 
 \subsubsection{\textbf{Baseline scheme}} The device keep sending the RACH request with the same power control threshold $\rho$.  
 
  \subsubsection{\textbf{Power ramping scheme}} 
\begin{figure}[t!]
        \centering
    \includegraphics[width=3.5in]{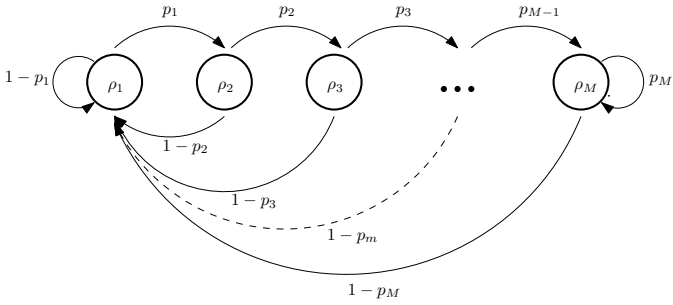}
    \caption{DTMC for the power ramping scheme for a device where each state represents the power control threshold used by the IoT element.}
\label{fig_Markov_power_ramping}
\end{figure}

The device increases its power control threshold in each RACH access attempt to increase the success probability until the maximum allowable threshold $\rho_M$ is reached. Let $\rho_m$ be the used power control threshold at the $m^{th}$ access attempt, then the power ramping strategy enforces $\rho_1 < \rho_2 <  \cdots < \rho_m < \cdots < \rho_M $. Upon RACH success, the device repeats the same strategy starting from the initial power control threshold $\rho_1$.  A schematic diagram for the device states in the power ramping scheme is shown in Fig.~\ref{fig_Markov_power_ramping}, where $p_m$ is the RACH access failure probability given that the device is using the power control threshold $\rho_m$.
  
  \subsubsection{\textbf{Back-off scheme}} 
\begin{figure}[t!]
        \centering
    \includegraphics[width=3.5in]{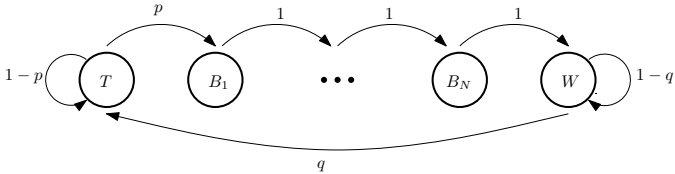}
        \caption{DTMC for the back-off scheme, where $T$ denotes the transmission state, $B_1, B_2,\cdots, B_N$ denote the deterministic back-off states, and $W$ denotes that random back-off state.}
\label{fig_Markov_backoff}
\end{figure}

The device goes for a deterministic back-off state for $N$ time slots followed by a probabilistic back-off state with probability $1-q$. The selected back-off scheme is general to capture deterministic back-off only by setting $q=1$, random back-off only by setting $N=0$, and generic combinations of both deterministic and random back-off states by setting $N>1$ and $q<1$. A schematic diagram for the device states in the back-off scheme is shown in Fig.~\ref{fig_Markov_backoff}, where $p$ is the RACH access failure probability.

It is worth mentioning that the baseline schemes is a special case of the power ramping scheme (i.e., by setting $M=1$) and also a special case of back-off scheme (i.e., by setting $N=0$ and $q=1$). Hence, the baseline scheme is used as a benchmark for both schemes to quantify the values of power ramping and transmission back-off on the network performance.


\subsection{Performance Metrics and Modeling Methodology} 

We consider two main performance metrics to assess the RACH access in LTE enabled IoT network, namely, the probability of RACH access failure in each time slot, denoted by $p$, and the expected waiting time for the RACH success, denoted by $ \mathcal{D}$.  Both performance metrics are functions of the received SINR at each transmission attempt. Specifically, the expected waiting time for RACH success can be expressed as 

\begin{equation}
\mathcal{D}= \frac{1}{(1-p) \mathcal{T}}
\end{equation}
\noindent where $\mathcal{T}$ is the probability that a device is transmitting on the RACH channel and $p$ is the probability of RACH transmission failure. The probability of  RACH transmission failure is given by
\begin{equation}
p=\mathbb{P}\left\{ \Upsilon(\tilde{\mathcal{U}}) < \theta \right\}
\end{equation}

\noindent where  $\Upsilon(\tilde{\mathcal{U}})$  is the SINR given that $\tilde{\mathcal{U}}$ device per unit area are transmitting on the same RACH ZC sequence, and $\theta$ is the SINR threshold defined for correct signal recovery. 

To assess the RACH performance, we first characterize the intra-cell and inter-cell interference on a test device using a typical ZC sequence. Both types of interference are characterized via the Laplace Transform (LT) of their probability density functions (PDFs). By virtue of the PPP and the uniform random selection of the ZC, all devices experience i.i.d interferences. Then, the obtained LTs are used to derive RACH transmission failure probability. The transmission probability for the baseline scheme is trivially equal to one. However, the transmission probabilities for each of the power ramping and back-off schemes should be obtained through the steady state probabilities of the DTMCs shown in Fig. \ref{fig_Markov_power_ramping} and Fig. \ref{fig_Markov_backoff}, respectively.  Note that both the power ramping scheme and the back-off scheme involve a causality problem. This is because the transmission probability $\mathcal{T}$ and RACH transmission failure probability $p$ are both unknown and are mutually dependent as can be inferred from (1) and (2) along with Fig. \ref{fig_Markov_power_ramping} and Fig. \ref{fig_Markov_backoff}. To overcome this causality problem we employ an iterative technique that involve the closed form expressions for the RACH transmission failure probability and the steady state probabilities of the DTMCs. 


\vspace{-1mm}

\section{Performance Analysis}

For the sake of organized presentation, we divide this section into three parts corresponding to each RACH access scheme.

\vspace{-3mm}
\subsection{Baseline Scheme}

The transmission probability of the baseline scheme is $\mathcal{T}=1$. The RACH transmission failure probability in the baseline scheme can be expressed in terms of the LT of the aggregate interferences as:

\begin{align}
p &=  \mathbb{P} \left\{ \frac{\rho h}{\sigma^2 + \mathcal{I}_{Intra}+\mathcal{I}_{Inter}}<\theta \right\}  \notag \\
& \overset{(a)}{=}1- \exp\left\{- \frac{\sigma^2 \theta}{\rho} \right\} \mathcal{L}_{\mathcal{I}_{Inter}} \left(\frac{\theta}{\rho} \right) \mathcal{L}_{\mathcal{I}_{Intra}} \left(\frac{\theta}{\rho} \right)
\label{SINR1}
\end{align}
where $h$ is the intended channel gain, ${\mathcal{I}_{Intra}}$ is the intra-cell interference,  ${\mathcal{I}_{Inter}}$ is the inter-cell interference, $\sigma^2$ is the noise power, and $\mathcal{L}_X(\cdot)$ denotes the LT of the PDF of $X$. Note that $(a)$ in \eqref{SINR1} follows from the exponential distribution of $h$ \cite{survey_h}. From \eqref{SINR1}, the RACH transmission failure probability is characterized via the following theorem

\begin{theorem}
\label{theorem1}
The {\rm RACH} transmission failure probability in the baseline scheme is given by \eqref{eq:Out1}, where ${}_2 F_1(.)$ is the Gauss hypergeometric function, $\mathcal{P}=\frac{c \lambda}{{c \lambda}+\tilde{\mathcal{U}}}$ and $c=3.575$ is a constant related to the approximate PDF of the PPP voronoi cell in $\mathbb{R}^2$.

\begin{figure*}
\begin{center}
\begin{align}\label{eq:Out1}
    {p} &\approx 1-  \left( \frac{\mathcal{P} (1+\theta)}{\mathcal{P}+\theta}\right)^{c} \exp\left\{- \frac{\sigma^2 \theta}{\rho} -  \frac{2 \theta \tilde{\mathcal{U}}}{(\eta-2) \lambda } \;{}_2F_1\left(1,1-2/\eta,2-2/\eta,-\theta\right) \right\} 
\end{align}
\end{center}
\hrulefill
\end{figure*}

\begin{proof}
See Appendix \ref{sec:AppA}.
\end{proof}
\end{theorem}

A special case of interest is for $\eta=4$, which is a typical path loss exponent for urban outdoor environment, is given 
by
\begin{align}
  {p}   &  {\approx} 1-  \left( \frac{\mathcal{P} (1+\theta)}{\mathcal{P}+\theta}\right)^{c} \exp\left\{- \frac{\sigma^2 \theta}{\rho} -  \frac{ \tilde{\mathcal{U}}  \sqrt{\theta}\arctan{\sqrt{\theta}} }{ \lambda }  \right\}
\label{eq:Out10}
\end{align}
The expression {in \eqref{eq:Out10}} gives the RACH transmission failure probability in terms of the elementary arctan function instead of the computationally complex Gauss hypergeometric function. 

From Theorem~\eqref{theorem1}, the average waiting time for RACH transmission success is given by $\mathcal{D} = \frac{1}{1-p}$.

\subsection{Power Ramping Scheme}

The power ramping scheme is more involved than the baseline scheme due to the state dependent RACH transmission failure probabilities along with the causality problem imposed by the dependence between the RACH transmission status (i.e., success or failure) and the interference. Let $\tilde{\Phi}_i \subseteq \tilde{\Phi}$ be the set of interfering devices transmitting with the power control threshold $\rho_i$, then the state dependent RACH transmission failure probability can be expressed as

\small
\begin{align}
p_m &=\mathbb{P} \left\{ \frac{\rho_m h}{\sigma^2 + \sum\limits_{i=1}^M  \mathcal{I}^{(i)}_{Intra}+   \sum\limits_{i=1}^M  \mathcal{I}^{(i)}_{Inter}}<\theta \right\}  \notag \\
  & \overset{(a)}{\approx}1- \exp\left\{- \frac{\sigma^2 \theta}{\rho_m} \right\} \prod\limits_{i=1}^{M} \mathcal{L}_{\mathcal{I}^{(i)}_{Inter}} \left(\frac{\theta}{\rho_m} \right) \mathcal{L}_{\mathcal{I}^{(i)}_{Intra}} \left(\frac{\theta}{\rho_m} \right)
 \label{SINR2}
\end{align}
\normalsize

\noindent where ${\mathcal{I}^{(i)}_{Intra}}$ and ${\mathcal{I}^{(i)}_{Inter}}$ are the intra-tier and inter-tier interferences from the devices in $\tilde{\Phi}_i$, respectively.  Note that $(a)$ in \eqref{SINR2} follows from the exponential distribution of $h$ and assuming that the point processes $\tilde{\Phi}_i$, $\forall i \in \{1,2,\cdots,M\}$ are mutually independent PPPs. The intensity of each point process $\tilde{\Phi}_i$ is given by $ \tilde{\mathcal{U}}_i = \mathbf{x}^{[i]} \tilde{\mathcal{U}}$, where $\mathbf{x}^{[i]}$ is the $i^{th}$ element in the steady state probability vector obtained from solving the power ramping DTMC in Fig.\ref{fig_Markov_power_ramping}. It is worth mentioning that the power ramping DTMC in Fig.\ref{fig_Markov_power_ramping} is irreducible, ergodic, and finite. Hence, the steady state probabilities exist~\cite{alfa}. 

The steady state probabilities of the power ramping  DTMC can be obtained by solving the system of linear equations 
\begin{equation}
\mathbf{x}\mathbf{P}=\mathbf{x}  \quad \text{and} \quad \mathbf{x}^T \mathbf{e} =1
\label{steady1}
\end{equation}
where $\mathbf{x}$ is the vector of the steady state probabilities, $\mathbf{e}$ is the vector of ones of length $M$, $(\cdot)^T$ denotes the transpose operator, and $\mathbf{P}$ is the steady state transmission matrix given by

\small
\begin{equation}
\mathbf{P} =  \begin{bmatrix}
1-p_1 & p_1 & 0 & 0 & \cdots  & 0\\ 
1-p_2 & 0 & p_2 & 0 & \cdots  & 0\\ 
1-p_3 & 0 & 0 & p_3 & \cdots  & 0\\ 
\vdots  & \vdots  & \vdots  & \vdots  & \ddots   & \vdots \\ 
1-p_{M-1} & 0 & 0 & 0 & 0 & p_{M-1}\\ 
1-p_M & 0 & 0 & 0 & 0 & p_M\\  
\end{bmatrix}
\label{powermat}
\end{equation}
\normalsize

Using \eqref{SINR2}, \eqref{steady1}, and \eqref{powermat}, the RACH transmission failure is characterized in the following theorem

\begin{theorem}
\label{theorem2}
For a given steady state probability vector $\mathbf{x}$, the average {\rm RACH} transmission failure probability in the power ramping scheme is given by
\begin{align}\label{outage_ramping}
 {p}=\displaystyle{\sum\limits_{m=1}^{M} \mathbf{x}^{[m]} p_m}
\end{align}
where $p_m$ is obtained by solving \eqref{eq:Out3}, $\mathcal{P}_k=\frac{c \lambda}{{c \lambda}+\tilde{\mathcal{U}}_k}$,  $\mathcal{U}_k= \mathbf{x}^{[k]} \tilde{\mathcal{U}}$, and $c=3.575$. The steady state probability $\mathbf{x}$ for the power ramping scheme is obtained via the following algorithm
\begin{figure*}
 \begin{center}
\begin{align}\label{eq:Out3}
    p_m  \approx 1- \exp\left\{- \frac{\sigma^2 \theta}{\rho_m} \right\} \prod\limits_{k=1}^{M}     \left( \frac{\mathcal{P}_k \left(1+\left(\frac{\theta\rho_k}{\rho_m} \right)\right)}{ \mathcal{P}_k +\left(\frac{\theta\rho_k}{\rho_m} \right)}\right)^{c} 
    \exp\Bigg\{  -  \frac{2 \tilde{\mathcal{U}}_k}{\lambda (\eta-2)} \; \left(\frac{\theta\rho_k}{\rho_m} \right)  \; _2 F_1\left(1,1-2/\eta,2-2/\eta,-\left(\frac{\theta\rho_k}{\rho_m} \right) \right) \Bigg\}, 
    \end{align}
    \hrulefill
    \begin{align}
    p_m  \overset{\eta=4}{\approx} 1- \exp\left\{- \frac{\sigma^2 \theta}{\rho_m} \right\} \prod\limits_{k=1}^{M}     \left( \frac{\mathcal{P}_k \left(1+\left(\frac{\theta\rho_k}{\rho_m} \right)\right)}{ \mathcal{P}_k +\left(\frac{\theta\rho_k}{\rho_m} \right)}\right)^{c}  \exp\left\{- \frac{\tilde{\mathcal{U}}_k}{\lambda} \sqrt{\frac{\theta\rho_k}{\rho_m}} \; \arctan{\left(\sqrt{\frac{\theta\rho_k}{\rho_m}} \right)} \right\}
    \label{eq:Out4}
\end{align}
 \end{center}
\hrulefill
 \end{figure*}
\begin{algorithm}[H]
Initialize $\boldsymbol{\rm x}(0)$ (e.g., with equiprobable states ).\\
\While { $\max|\boldsymbol{\rm x}(i) - \boldsymbol{\rm x}(i-1) |\geq \epsilon $  } {
			1- Calculate $p_m$, $\forall m \in \{1,2,\cdots,M\}$ using $\boldsymbol{x}(i-1)$ and \eqref{eq:Out3}.\\
  				2- Update  $\boldsymbol{\rm P}$ in \eqref{powermat} with $p_m$, $\forall m \in \{1,2,\cdots,M\}$.  \\
3-   Obtain $\boldsymbol{\rm x}(i)$ by solving  \eqref{steady1}.
 				}\;  	
				return $\boldsymbol{\rm x} \leftarrow \boldsymbol{\rm x}(i)$.		
 \caption{Computation of $\boldsymbol{\rm x}$ for the power ramping scheme}
\end{algorithm}
\begin{proof}
See Appendix \ref{sec:AppB}.
\end{proof}
\end{theorem}

A special case of interest is for $\eta=4$, which is a typical path loss exponent for urban outdoor environment, is given 
by \eqref{eq:Out4} where the Gauss hypergeometric  function reduces to the elementary arctan function. 

From Theorem~\eqref{theorem2}, the average waiting time for RACH transmission success is given by $\mathcal{D} = \frac{1}{1-p}$.

\subsection{Back-off Scheme}

Different from the baseline and power ramping scheme, the back-off scheme is a more conservative transmission scheme, where some of the devices stop transmission and go into back-off states to relief the congestion on the RACH channel. From the back-off DTMC shown in Fig.~\ref{fig_Markov_backoff}, it is clear that only the devices in state $T$ are the transmitting devices. Without loss of generality, we let the $T$ to be the first element in the steady state probability vector $\mathbf{x}$ followed by the $N$ back-off slots and finally the probabilistic back-off state. Hence, the intensity of simultaneously active devices is $\mathbf{x}^{[0]} \tilde{\mathcal{U}}$.  The RACH transmission failure probability in the back-off scheme can be expressed as

{
\small
\begin{align}
p =1- \exp\left\{- \frac{\sigma^2 \theta}{\rho} \right\} \mathcal{L}_{\mathcal{I}_{Inter}} \left(\frac{\theta}{\rho} \right) \mathcal{L}_{\mathcal{I}_{Intra}} \left(\frac{\theta}{\rho} \right)
 \label{SINR3}
\end{align}
\normalsize
}
\noindent where ${\mathcal{I}_{Intra}}$ and ${\mathcal{I}_{Inter}}$ are the intra-tier and inter-tier interferences from the devices in the transmission state (i.e., not in back-off). The probability of being in the transmission mode  $\mathbf{x}^{[0]}$ is obtained from solving the back-off DTMC in Fig.~\ref{fig_Markov_backoff}, which is irreducible, ergodic, and finite. Hence, its steady state probabilities exist~\cite{alfa}.

The steady state probabilities of back-off DTMC can be obtained by solving the system of linear equations in  \eqref{steady1} but with the following probability transition matrix 

\small
\begin{equation}
\mathbf{P} = \begin{bmatrix}
1-p & p & 0 & 0 & \cdots  & 0\\ 
0 & 0 & 1 & 0 & \cdots  & 0\\ 
0 & 0 & 0 & 1 & \cdots  & 0\\ 
\vdots  & \vdots  & \vdots  & \vdots  & \ddots   & \vdots \\ 
0 & 0 & 0 & 0 & 0 & 1 \\ 
q & 0 & 0 & 0 & 0 & 1-q \\ 
\end{bmatrix}
\label{power3}
\end{equation}
\normalsize

Using \eqref{steady1}, \eqref{SINR3}, and \eqref{power3}, the RACH transmission failure probability is characterized in the following theorem

\begin{theorem}
\label{theorem3}
For a given steady state probability vector $\mathbf{x}$, the average {\rm RACH} transmission failure probability in the back-off scheme is given by \eqref{eq:Out5},  where $\mathcal{P}_B=\frac{c \lambda}{{c \lambda}+\tilde{\mathbf{x}^{[0]} \mathcal{U}}}$,  $\mathcal{U}_B= \mathbf{x}^{[0]} \tilde{\mathcal{U}}$, and $c=3.575$. The steady state probability $\mathbf{x}$ for the back-off scheme is obtained via the following algorithm
\begin{figure*}
\begin{center}
\begin{align}\label{eq:Out5}
    {p} &\approx 1-  \left( \frac{\mathcal{P}_B (1+\theta)}{\mathcal{P}_B+\theta}\right)^{c} \exp\left\{- \frac{\sigma^2 \theta}{\rho} -  \frac{2 \theta \tilde{\mathcal{U}}_B}{(\eta-2) \lambda } \;{}_2F_1\left(1,1-2/\eta,2-2/\eta,-\theta\right) \right\} 
\end{align}
\end{center}
\hrulefill
\end{figure*}
\begin{algorithm}[H]
Initialize $\boldsymbol{\rm x}(0)$ (e.g., $\boldsymbol{\rm x}^{[0]}=1$ ).\\
\While { $\max|\boldsymbol{\rm x}(i) - \boldsymbol{\rm x}(i-1) |\geq \epsilon $  } {
			1- Calculate $p$ using $\boldsymbol{x}(i-1)$ and \eqref{eq:Out5}.\\
  				2- Update  $\boldsymbol{\rm P}$ in \eqref{power3} with $p$.  \\
3-   Obtain $\boldsymbol{\rm x}(i)$ by solving  \eqref{steady1}.
 				}\;  	
				return $\boldsymbol{\rm x} \leftarrow \boldsymbol{\rm x}(i)$.		
 \caption{Computation of $\boldsymbol{\rm x}$  for the back-off scheme}
\end{algorithm}
\begin{proof}
Similar to Theorem 1 and Theorem 2.
\end{proof}
\end{theorem}

A special case of interest is for $\eta=4$, which is a typical path loss exponent for urban outdoor environment, is given 
by

\small
\begin{align}
  {p}   &  {\approx} 1-  \left( \frac{\mathcal{P}_B (1+\theta)}{\mathcal{P}_B+\theta}\right)^{c} \exp\left\{- \frac{\sigma^2 \theta}{\rho} -  \frac{ \tilde{\mathcal{U}}_B  \sqrt{\theta}\arctan{\sqrt{\theta}} }{ \lambda }  \right\}
\label{eq:Out2}
\end{align}
\normalsize 
 
 From Theorem~\eqref{theorem3}, the average waiting time for RACH transmission success for the back-off scheme is given by $\mathcal{D} = \frac{1}{\mathbf{x}^{[0]} (1-p)}$.
\vspace{1 mm}

The design parameters $N$ and $q$ in the back-off scheme impose a trade off between the interference level and transmission probability on the RACH. Selecting a large $N$ or small $q$ lead to a conservative spectrum access (i.e., low $\mathbf{x}^{[0]}$) with high transmission success probability (i.e., high $(1-p)$), and vice versa. Hence, the selection of $N$ and $q$ are crucial to balance this tradeoff. The optimal values of $N$ and $q$ can be obtained by solving the following optimization problem 

\begin{equation}
\begin{aligned}
& \underset{N,q}{\text{minimize}}
& & \mathcal{D} = \frac{1}{\mathbf{x}^{[0]} (1-p)} \\
& \text{subject to}
& & N \in \mathbb{Z} \\
& & & 0 \leq q \leq 1 
\end{aligned}
\label{optimization}
\end{equation}
Due to space constraints, techniques for solving \eqref{optimization} is out of the scope of this paper. Hence, \eqref{optimization} is solved via exhaustive search. 
\vspace{-0.4 mm}

\section{Numerical Results \& Simulations}\label{sec:Results}
\begin{figure}[t!]
\centering
\includegraphics[width=3.3in]{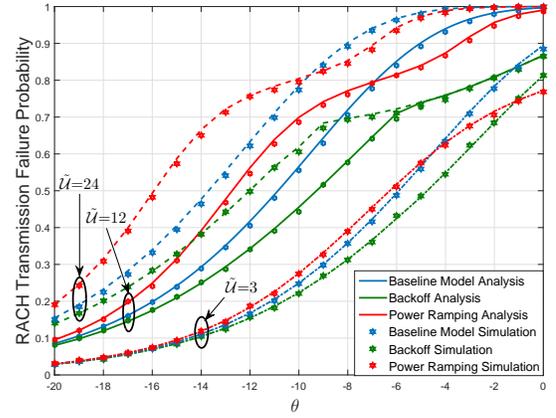}
\caption{Outage Probability of the three schemes.}\label{fig:outage}
\end{figure}

\begin{figure}[t!]
\label{fig:Local_delay}
\centering
\includegraphics[width=3.3in]{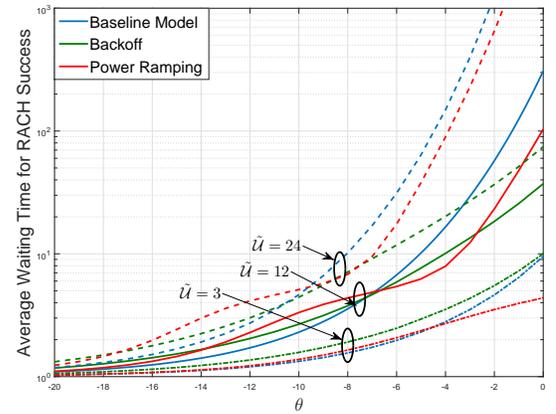}
\caption{Local Delay of the three schemes}\label{fig:delay}
\end{figure}

At first we compare the proposed analysis with independent system level simulations for the baseline, power ramping, and back-off schemes. The values of $N$ and $q$ in the back-off scheme are obtained via \eqref{optimization} for every $\theta$. In the  power ramping scheme, the values of  $\rho_m$ are chosen to vary from $-90$ dBm to $-70$ dBm with $4$ dBm resolution. In the simulation scenarios we consider a PPP cellular network over a 100 km$^2$ area. Devices are distributed according to an independent PPP and each device associates to its closest BS. The devices employ channel inversion power control. The collected statistics are taken for devices located within 1 km from the origin to avoid the edge effects. Unless otherwise stated, we choose { $\tilde{\mathcal{U}} =3,\; 12,\; \text{and}\; 24$} device/km$^2$/ZC-sequence, $\lambda = 3$  BS/km$^2$ , $\eta =4$, $\rho =-90$ dBm, $\sigma^2 =-90$ dBm, and $ -20 \leq \theta \leq 0$ dB. Fig.~\ref{fig:outage} depicts the results for the RACH transmission failure probability while Fig.~\ref{fig:delay} shows the average waiting time for RACH transmission success for each scheme. It is worthwhile to note first the close fit between analysis and simulation as per Fig.~\ref{fig:outage}, which validates the developed mathematical framework.

For insightful conclusions,  Fig.~\ref{fig:outage} and Fig.~\ref{fig:delay} should be considered jointly. While one scheme may be favorable from the RACH transmission failure probability perspective, it may be invoking too much delay to attain such a good RACH transmission failure probability, and hence, shall have adverse impact on the average waiting time for RACH transmission success. As such, there are three key takeaway messages from the presented results. At low SINR thresholds and/or low intensity, it is better to transmit persistently whenever a packet exists. The power ramping scheme only results in excessive interference in the network therefore increasing the RACH transmission failure probability. For moderate SINR values and/or device intensities, it is actually better to ramp up power so as to prioritize devices which already experienced RACH transmission failures. In comparison, back-off imposes unnecessary delays that increase  the average waiting time for the RACH transmission success. Finally, at high SINR region and/or intensities, the interference obviously becomes overwhelming. Accordingly, this is the effective scenario for the back-off scheme where devices should take some idle time before reattempting to access in order to relief RACH congestion. Consequently, interference levels are relaxed, RACH transmission failure probability is reduced while providing a better average waiting time for RACH transmission success.
 
  The optimum values for the back-off parameters are reported in Table \ref{Modes_Table}. A zero value for $N$ means that a RACH failure is immediately followed by a transition to a random waiting state, as shown in Fig. \ref{fig_Markov_backoff}. Furthermore, $q=1$ implies an immediate transitions to the transmit state rather than back-off. It is obvious from Table \ref{Modes_Table} that at low device intensity, the combination $N=0$ and $q=1$ prevails regardless of the threshold $\theta$. The intuition is that with a low number of connections, persistent transmission attempts is simply the best strategy. As the device intensity grows, persistent transmission is only effective at low SINR thresholds. On the other hand, when the device intensity and $\theta$ grow, the optimum values for $N$ and $q$ increases and decreases, respectively, which implies that back-off becomes crucial for network operation. In other words, devices need to back-off for at least $N$ time slots followed by a randomized spectrum coordination with probability $q$ before they attempt to transmit again. This is necessary to relief the congestion  and reduce the interference in the network to maintain an acceptable RACH transmission success probability and waiting time. 
  
  An interesting observation from Figure~\ref{fig:outage} and Table \ref{Modes_Table} is that at high device intensity (i.e., $\tilde{\mathcal{U}}=24$) and SINR thresholds, the optimal values of $N$ and $q$ maintains a similar RACH transmission failure probability to the moderate intensity scheme (i.e., $\tilde{\mathcal{U}}=12$). This behavior offloads the access delay from the uncontrollable transmission failures to the controllable back-off coordination/contention order to minimize the overall average waiting time for RACH success.  
 
 Despite that the random back-off scheme outperforms the baseline and power ramping schemes at high intensities, Fig.~\ref{fig:delay}  shows significant degradation in the average waiting time for the back-off scheme as the intensity increases, which confirms the vulnerability of the LTE RACH to support the massive IoT traffic. Hence, novel scalable solutions for RACH access and resource allocations in LTE enabled IoT networks are required, which will be the focus of our future work.



\begin{table}[h!]
\centering
\small
\caption{{\color{black} Optimum Values for Back-off Parameters } }
\resizebox{0.45 \textwidth}{!}{\begin{tabular}{|c|c|c|c|}
\hline
\rowcolor{lightgray}\multirow{-2}{*}{}
& & &  \\

\rowcolor{lightgray}\multirow{-2}{*}{Device intensity (\textbf{$\tilde{\mathcal{U}}$})}&  \multirow{-2}{*}{SINR threshold (\textbf{$\theta$})}&  \multirow{-2}{*}{\# back-off slots (\textbf{$N$})}&  \multirow{-2}{*}{ back-off probability  (\textbf{$q$})}  \\ \hline \hline

 \multirow{6}{*}{\textbf{$3 $\; UEs / km$^2$}} & \multirow{2}{*}{$-10$ dB} &  \multirow{2}{*}{$ 0  $} & \multirow{2}{*}{$1$}  \\
 & &  & \\ \cline{2-4}
 & \multirow{2}{*}{$-6$ dB} & \multirow{2}{*}{$ 0 $} &   \multirow{2}{*}{$ 1 $}  \\
& & & \\  \cline{2-4}
 & \multirow{2}{*}{$-2$ dB} &  \multirow{2}{*}{$ 0 $}  & \multirow{2}{*}{$1 $}  \\
 & & &  \\   \hline
  \hline

  \multirow{6}{*}{\textbf{$12 $\; UEs / km$^2$}} & \multirow{2}{*}{$-10$ dB} &  \multirow{2}{*}{$ 0  $} & \multirow{2}{*}{$1$}  \\
 & &  & \\ \cline{2-4}
 & \multirow{2}{*}{$-6$ dB} & \multirow{2}{*}{$ 0 $} &   \multirow{2}{*}{$ 1 $}  \\
& & & \\  \cline{2-4}
 & \multirow{2}{*}{$-2$ dB} &  \multirow{2}{*}{$ 2 $}  & \multirow{2}{*}{$ .91 $}  \\
 & & &  \\   \hline
  \hline

  \multirow{6}{*}{\textbf{$24 $\; UEs / km$^2$}} & \multirow{2}{*}{$-10$ dB} &  \multirow{2}{*}{$ 0  $} & \multirow{2}{*}{$1$}  \\
 & &  & \\ \cline{2-4}
 & \multirow{2}{*}{$-6$ dB} & \multirow{2}{*}{$ 2 $} &   \multirow{2}{*}{$ .87 $}  \\
& & & \\  \cline{2-4}
 & \multirow{2}{*}{$-2$ dB} &  \multirow{2}{*}{$ 6 $}  & \multirow{2}{*}{$.69 $}  \\
 & & &  \\   \hline
\end{tabular}}
\label{Modes_Table}
\end{table}

\section{Conclusions}

This paper introduces a tractable stochastic geometry model to study random access in LTE networks. The prime focus is to apply the model in the context of IoT where massive numbers of devices with accordingly huge traffic volumes need to be accommodated. The model is used to assess and contrast three key access schemes, namely baseline, power ramping, and back-off schemes. Closed form expressions for the RACH transmission success probability are obtained in terms of the devices states, in which the device states are obtained by solving scheme dependent discrete time Markov chains. The results clearly suggest that persistent transmission (i.e. baseline scheme) is preferable at relatively low device intensities and/or low detection thresholds. As the intensity and/or threshold grows, it is shown that the power ramping scheme becomes favorable to be then knocked out by the back-off scheme, which offers the lowest local delay at high intensities and/or detection thresholds. For instance, at $0$ dB SINR detection threshold, $64$ orthogonal sequences per BS, and 3 BSs/km$^2$, the back-off scheme offers $65\%$ and $99\%$ RACH access delay reduction when compared to the power ramping scheme for $768$ and $1536$ devices/km$^2$ (i.e., $256$ and $512$ devices/BS), respectively. 


\appendix

\subsection{Proof of Theorem 1}\label{sec:AppA}
To evaluate the interference experienced by a device, we find the Laplace Transform(LT) of the aggregate intra-cell interference along with the inter-cell interference. Note that the nearest BS association and the employed power control enforce the following two conditions; (i) the intra-cell interference from an interfering device is equal to $\rho$, and (ii) the inter-cell interference from any interfering device is strictly less that $\rho$. Approximating the set of interfering devices by a PPP with independent transmit powers, the aggregate inter-cell interference received at the BS is obtained as:

\small
\begin{align}\label{eq:AppA_3}
\mathcal{I}_{Inter}=  \sum\limits_{u_i\in \Phi \setminus \{o\} } \mathbbm{1}_{\{P_{i} \parallel u_i\parallel^{-\eta}<\rho\}}P_{i} h_i \parallel u_i\parallel^{-\eta} 
\end{align}
\normalsize
The Laplace Transform of \eqref{eq:AppA_3} is obtained as:
\small
\begin{align}
&\mathcal{L}_{\mathcal{I}_{Inter}}(s) 
&=\exp\left(-2\pi \; \tilde{\mathcal{U}} \; s^{\frac{2}{\eta}} \;\mathbb{E}_{P}\left[P^{\frac{2}{\eta}} \;   \right] \int\limits_{(s\rho)^{\frac{-1}{\eta}}}^{\infty} \frac{y}{y^\eta +1} dy  \right).
\end{align}
\normalsize

\noindent  The LT is obtained by using the probability generating function (PGFL) of the PPP \cite{ martin_book} and following \cite{elsawy2014stochastic}, where  $\mathbb{E}_{x} [.]$ is the expectation with respect to the random variable $x$ , the Laplace Transform is obtained by substituting the value of $\mathbb{E}_{P}\left[P^{\frac{2}{\eta}}\right]$ from [Lemma 1,\cite{elsawy2014stochastic}].

The Intra-cell interference conditioned on the number of neighbors is given by:
\begin{align}\label{eq:AppA_1}
\mathcal{I}_{Intra\mid N} = \displaystyle{\sum\limits_{n=1}^{N} \rho h}
\end{align}
The Laplace Transform of \eqref{eq:AppA_1} is obtained as:
\begin{align}
\mathcal{L}_{\mathcal{I}_{Intra}\mid N}(s)&= \mathbb{E}[e^{-s\mathcal{I}}]\notag\\
										&= \frac{1}{(1+s\rho)^N}
\end{align}

By considering that there is only Inter-cell interference when the number of neighbors in the cell is 0, and both of inter-cell and intra-cell interference otherwise, and deconditioning on the distribution  $\mathcal{N}$ which is found in \cite{6576413} we can write equation~(\ref{SINR1}) as follows:

\footnotesize
\begin{align}\label{eq:AppA_2}
1- \exp\left\{- \frac{\sigma^2 \theta}{\rho} \right\} \mathcal{L}_{\mathcal{I}_{Inter}} \left(\frac{\theta}{\rho} \right)\left[ \mathbb{P} \left\{\mathcal{N}=0 \right\}+{\sum\limits_{n=1}^{\infty}\frac {\mathbb{P} \left\{\mathcal{N}=n \right\}}{(1+s\rho)^n}} \right]
\end{align}
\normalsize

after some manipulations, \eqref{eq:Out1} in Theorem 1 is obtained.

\vspace{0.1cm}
\subsection{Proof of Theorem 2}\label{sec:AppB}
The intra-cell interference in this case is:
\begin{align}
\mathcal{I}_{Intra}=\displaystyle{ \sum\limits_{k=1}^{M}\sum\limits_{n=1}^{N} \rho_k h}
\end{align}
 while the inter-cell interference is:
\begin{align}
\mathcal{I}_{Inter}=\displaystyle{ \sum\limits_{k=1}^{M} \; \sum\limits_{u_i\in \Phi \setminus \{o\}} \mathbbm{1}_{\{P_{ik} \parallel u_i\parallel^{-\eta}<\rho_k\}}P_{ik} h_i \parallel u_i\parallel^{-\eta} }
\end{align}
using the same procedure in the proof of Theorem 1, and the total probability theorem, \eqref{outage_ramping} is obtained.
\vspace{0.001cm}
\section*{Acknowledgement}
The authors would like to acknowledge KAUST for funding this work, and Dr. Abdulkareem Adinoyi for his valued comments and suggestions.
\vspace{0.001cm}

\bibliographystyle{IEEEtran}
\bibliography{RACH}
\vfill
\end{document}